\documentclass[12pt]{article}
\usepackage[english,german,french,polish]{babel}
\usepackage[T1]{fontenc}
\usepackage{amsfonts}

\begin{document}

\selectlanguage{english}

\baselineskip 0.73cm
\topmargin -0.4in
\oddsidemargin -0.1in

\let\ni=\noindent

\renewcommand{\thefootnote}{\fnsymbol{footnote}}

\newcommand{\SM}{Standard Model }

\pagestyle {plain}

\setcounter{page}{1}



~~~~~~
\pagestyle{empty}

\begin{flushright}
IFT-- 09/6
\end{flushright}

\vspace{0.4cm}

{\large\centerline{\bf Photonic portal to the hidden sector}}

{\large\centerline{\bf and the electroweak symmetry}}

\vspace{0.5cm}

{\centerline {\sc Wojciech Kr\'{o}likowski}}

\vspace{0.3cm}

{\centerline {\it Institute of Theoretical Physics, University of Warsaw}}

{\centerline {\it Ho\.{z}a 69, 00--681 Warszawa, ~Poland}}

\vspace{0.6cm}

{\centerline{\bf Abstract}}

\vspace{0.2cm}

\begin{small}


A weak photon interaction with the hidden sector of the Universe, introduced recently to realize a "photonic portal"\, 
(to such a hypothetic sector responsible for cold dark matter), is conjectured to be embedded in a more extended weak 
interaction displaying electroweak symmetry spontaneously broken by the Standard-Model Higgs mechanism. This
is a hypothetic new weak interaction between hidden and Standard-Model sectors of the Universe, appearing in our 
model in addition to the conventional electroweak interaction acting in the Standard-Model sector.
 
\vspace{0.6cm}

\ni PACS numbers: 14.80.-j , 04.50.+h , 95.35.+d 

\vspace{0.3cm}

\ni May 2009

 
\end{small}

\vfill\eject

\pagestyle {plain}

\setcounter{page}{1}

\vspace{0.5cm}

\ni {\bf 1. Introduction}

\vspace{0.4cm} 

In a recent work [1], we discussed a model of a hidden sector of the Universe consisting of sterile spin-1/2 fermions ("sterinos") and sterile spin-0 bosons ("sterons") interacting weakly through the mediation of sterile quanta of an 
antisymmetric-tensor field $A_{\mu\,\nu}$ ("$\!A$ bosons"), but with a stronger coupling than through the universal gravity. It is assumed that, after the electroweak symmetry $SU(2)\times U(1)$ of \SM is spontaneously broken by the Standard-Model Higgs mechanism, these sterile particles of hidden sector can communicate with  the Standard-Model sector through the electromagnetic field $ F_{\mu\,\nu} = \partial_\mu A_\nu - \partial_\nu A_\mu $ weakly coupled (in pair with the steron field $\varphi$) to the $A$ boson field $A_{\mu\,\nu}$: 

\begin{equation}
-\frac{1}{2} \sqrt{f} \varphi F_{\mu \nu} A^{\mu\,\nu} 
\end{equation}

\ni ("photonic portal"\, to the hidden sector responsible for cold dark matter). Here, the field $A_{\mu\,\nu}$ is nongauge and has dimension one, in contrast to the gauge field $F_{\mu\,\nu}$ of dimension two. It provides two kinds of spin-1 quanta with parity $-$ and + , expected to define a large mass scale $M$. Also the sterino field $\psi$ is assumed to be coupled to the sterile mediating field $A_{\mu\,\nu}$ :

\begin{equation}
-\frac{1}{2} \sqrt{f\,} \zeta \bar\psi \sigma_{\mu\,\nu} \psi  A^{\mu \nu}\,.
\end{equation}

\ni In Eqs. (1) and (2), $\sqrt{f}$ and $\sqrt{f\,} \zeta $ are two dimensionless coupling constants ($f > 0$), expected to be small, so that the interactions (1) and (2) are considered as weak.

We can see that the electromagnetic field ($A_\mu $ or $ F_{\mu\,\nu} = \partial_\mu A_\nu - \partial_\nu A_\mu $) participates here in two different interactions: the Standard-Model electromagnetic interaction $-j_\mu A^\mu $ with $j_\mu$ denoting the Standard-Model electric current and a new weak nongauge interaction (1) (both are electromagnetically gauge invariant, the first due to the electric charge conservation $\partial_\mu j^\mu = 0$, the second explicitly because of the dependence on $F_{\mu\,\nu}$.

Note that the phenomenon of different interactions for the same particles is rather typical in particle physics, leading usually to the interaction unification. For instance, leptons and quarks display, in addition to universal gravity, two and three phenomenologically different interactions, respectively, unified at high energies into spontaneously broken electroweak symmetry of \SM and, possibly, into such symmetries of GUTs.

For the steron field $\varphi$ appearing, in particular, in the interaction (1), the expansion

\begin{equation} 
\varphi = <\!\!\varphi\!\!>_{\rm vac} + \,\varphi_{\rm ph}
\end{equation}

\ni is conjectured, where $<\!\!\varphi\!\!>_{\rm vac} \neq 0$ and $\varphi_{\rm ph}$ are a spontaneously nonzero vacuum expectation value of $\varphi$ and the physical steron field, respectively. The value  $<\!\!\varphi\!\!>_{\rm vac} \neq 0$ can be used to generate spontaneously masses of all sterile particles: sterinos, sterons and $A$ bosons [1].

Due to the interactions (1) and (2), the fields $F_{\mu\,\nu}$ and $A_{\mu\,\nu}$ satisfy in the presence of hidden sector the "\,$\!$supplemented Maxwell's equations"

\begin{equation}
\partial^\nu (F_{\mu\,\nu} +  \sqrt{f\,} \varphi A_{\mu\,\nu}) = -j_\mu \;,\; F_{\mu\,\nu} = \partial_\mu A_\nu - \partial_\nu A_\mu 
\end{equation}

\ni and

\begin{equation}
(\Box - M^2)A_{\mu\,\nu} = - \sqrt{f\,} (\varphi F_{\mu\,\nu} + \zeta \bar\psi \sigma_{\mu\,\nu} \psi)\;. 
\end{equation}

In the case of small momentum transfers {\it versus} a large mass scale $M$, the field equation (4) and interaction (1) imply approximately the effective Fermi-like coupling (involving the field pairs $\varphi F_{\mu\,\nu}$ and $\bar{\psi}\sigma_{\mu\,\nu}\psi$): 

\begin{equation}
  -\frac{1}{4} \frac{f}{M^2}\left(\varphi F_{\mu \nu} + \zeta \bar\psi \sigma_{\mu\,\nu} \psi \right) \left(\varphi F^{\mu \nu} + \zeta \bar\psi \sigma^{\mu\,\nu} \psi \right) \,,
\end{equation}

\ni where its cross term (with $\varphi = <\!\!\varphi\!\!>_{\rm vac}$)

\begin{equation}
- \frac{f \zeta <\!\!\varphi\!\!>_{\rm vac}}{2M^2} \bar\psi \sigma_{\mu\,\nu} \psi F^{\mu \nu}  
\end{equation}

\ni shows that sterinos get a small  magnetic moment


\begin{equation}
 \mu_\psi \equiv \frac{f \zeta<\!\!\varphi\!\!>_{\rm vac}}{2M^2} 
\end{equation}

\ni spontaneously generated by $<\!\!\varphi\!\!>_{\rm vac} \neq 0$. Of course, they remain electrically neutral, as the correction 

\begin{equation}
\delta j_\mu \equiv \partial^\nu(\sqrt{f\,} \varphi A_{\mu \nu}) 
\end{equation}

\ni to the electric current in Eq. (4) gives a zero correction 

\vspace{-0.2cm}

\begin{equation}  
\delta Q \equiv \int\! d^{\,3}x \delta j_0 = \int d^{\,3}x \partial^{\,l}(\sqrt{f\,} \varphi A_{0 l} ) = 0
\end{equation}

\ni to the electric charge $Q = \int\! d^3x j_0$.

We should like to observe from Eqs. (3) and (6) that due to $<\!\!\varphi\!\!>_{\rm vac} \!\neq\! 0$ a phenomenon of finite additional renormalization ("primordial renormalization") [1] appears in this model.  

It is important to note that the interaction (1), working after the spontaneous breakdown of electroweak symmetry, breaks by itself this symmetry explicitly if it is introduced to the total Lagrangian from the very beginning (before the symmetry breakdown). This may not be a satisfactory feature of our model. Therefore, in the next Section, we will try to embed our weak photon interaction (1) in a more extended form, preserving the electroweak symmetry before it is spontaneously broken by the Standard-Model Higgs mechanism. In consequence of this reconstruction, however, our model shall change a lot. In particular, in place of steron field $\varphi$ there will appear four scalar fields $ \varphi^+, \varphi^-,\varphi^0$ and $\varphi$ realizing a reduced quartet $(3+1)\times 1$ of the electroweak symmetry $SU(2)\times U(1)$ (a scalar analogy of $W^+_\mu, W^-_\mu, W^0_\mu$ and $B_\mu$).     

\vspace{0.4cm}

\ni {\bf 2. "Photonic portal"\, embedded in the electroweak symmetry} 

\vspace{0.4cm}

Now, in addition to the weak sterino interaction (2), let us consider (in place of the weak photon coupling (1)) the following more extended form preserving the electroweak symmetry before it is spontaneously broken by the Standard-Model Higgs mechanism:

\begin{equation}
 -\frac{1}{2}\left( \sqrt{f\,}\sum_i\varphi_i W_i^{\mu \nu} + \sqrt{f'\,} \varphi B^{\mu\,\nu} \right) A_{\mu \nu} \,,
\end{equation}

\vspace{-0.2cm}

\ni where

\vspace{-0.2cm}

\begin{equation}
W_i^{\mu \nu} = \partial^\mu W^\nu_i - \partial^\nu W^\mu_i + g \sum_{j\,k} \varepsilon_{i\,j\,k} W_j^{\mu } W_k^{\nu }\; \,,\;\, B^{\mu \nu} = \partial^\mu B^\nu - \partial^\nu B^\mu 
\end{equation}

\ni and, after spontaneously breaking the electroweak symmetry,

\vspace{-0.2cm}

\begin{eqnarray} 
W^\mu_1   \!=\!  \frac{1}{\sqrt{2}} \left(W^{+\mu} + W^{\!-\!\mu}\right)\,,\, W^\mu_2\!\!\!  & \!=\! & \!\!\!\frac{1}{i\sqrt{2}} \left(W^{+\mu} \!-\! W^{-\mu}\right) \,,\, W^\mu_3   \!=\!  W^{0 \mu} \!=\! \cos\theta_w Z^{\mu} \!-\! \sin\theta_w A^{\mu}\,, \nonumber \\ B^\mu & \!=\! & \sin\theta_w Z^{\mu} \!+\! \cos\theta_w A^{\mu}\,, 
\end{eqnarray}

\vspace{-0.2cm}

\ni while 

\vspace{-0.2cm}

\begin{equation}
\varphi_1 = \frac{1}{\sqrt{2}} \left(\varphi^+ + \varphi^-\right)\,,\, \varphi_2 = \frac{1}{i\sqrt{2}} \left(\varphi^+ - \varphi^-\right)\,,\, \varphi_3 = \varphi^0 
\end{equation}

\ni and $\varphi$  are four scalar fields transforming under the electroweak symmetry $SU(2)\times U(1)$ as a reduced quartet $(3+1)\times 1$.  Here, $e = g \sin\theta_w = g' \cos\theta_w >0$, whereas $\sqrt{f\,}$ and $\sqrt{f'\,}$ are two dimensionless coupling constants ($f>0$ and $f'>0$) presumed to be small.

Then, inserting Eqs. (13) and (14) into the interaction (11), we obtain after the spontaneous breaking of electroweak symmetry that

\begin{eqnarray} 
-\frac{1}{2}\!\left(\!\! \sqrt{\!f}\sum_i\!\!\varphi_i\right.\!\!\!\!\! &W_i^{\mu \nu}\!\! & \!\!\!\! + \left.\sqrt{\!f'}\varphi B^{\mu \nu}\!\right) \!A_{\mu \nu} \nonumber \\ & \!\!=\!\! &\!\!\!\!\! -\frac{1}{2}  \sqrt{\!f}  \varphi^- \left[\partial^\mu W^{+ \nu} -  \partial^\nu W^{+ \mu}- i g \left(W^{+\mu} W^\nu_3 - W^\mu_3 W^{+\nu} \right)\!\right] A_{\mu \nu} \nonumber \\ & \!\!\!\! &\!\!\!\!\! -\frac{1}{2}\sqrt{\!f} \varphi^+ \left[\partial^\mu W^{- \nu} -  \partial^\nu W^{-\mu} +  i g \left(W^{-\mu} W^\nu_3 - W^\mu_3W^{-\nu}\right)\!\right] A_{\mu \nu} \nonumber \\ & \!\!\!\! &\!\!\!\!\! -\frac{1}{2}\!\left[\!\sqrt{\!f^{(Z)}} \varphi^{(Z)} Z^{\mu \nu} \!\!+\! \sqrt{\!f^{(A)\,}}\varphi^{(A)}F^{\mu \nu}\!\!+\! \sqrt{\!f} ig \varphi_3\! \left(\!W^{+\mu}W^{-\nu} \!\!\!-\! W^{-\mu}W^{+\nu} \!\right)\!\right] \!A_{\mu \nu} , \nonumber \\
\end{eqnarray}

\ni where 

\begin{equation} 
Z^{\mu \nu}  = \partial^\mu Z^{\nu} - \partial^\nu Z^{\mu} \;\;,\;\; F^{\mu \nu} = \partial^\mu A^{\nu} \!-\!  \partial^\nu A^{ \mu}
\end{equation}

\ni and 

\begin{equation} 
f^{(Z)} \equiv f \cos^2\theta_w + f' \sin^2\theta_w \;\,,\;\, f^{(A)} \equiv f \sin^2\theta_w + f' \cos^2\theta_w \,,
\end{equation}

\ni while

\vspace{-0.2cm}

\begin{eqnarray}
\sqrt{\!f^{(Z)}} \varphi^{(Z)} & \equiv & \sqrt{\!f} \cos\theta_w \varphi_3 + \sqrt{\!f' }\sin\theta_w \varphi \,, \nonumber \\
\sqrt{\!f^{(A)}} \varphi^{(A)} & \equiv & \!\!\!-\sqrt{\!f} \sin\theta_w \varphi_3 + \sqrt{\!f' }\cos\theta_w \varphi \,.
\end{eqnarray}

If it happens that $f = f'$, then $f^{(Z)} = f^{(A)} = f = f'$ and

\begin{equation} 
\varphi^{(Z)} \equiv \cos\theta_w\varphi_3 + \sin\theta_w \varphi \;\,,\,\; \varphi^{(A)} \equiv -\sin\theta_w \varphi_3 + \cos\theta_w \varphi
\end{equation}

\ni are the same linear combinations as those for $W^\mu_3$ and $B^\mu $   in the Standard-Model mixing

\begin{equation} 
Z^\mu = \cos\theta_w W^\mu_3 + \sin\theta_w B^\mu \;\,,\,\; A^\mu = -\sin\theta_w W^\mu_3 + \cos\theta_w B^\mu \,.
\end{equation}

\ni In this case, the conjecture

\begin{equation}
f = f' = e^2
\end{equation}

\ni with $e^2 =  4\pi \alpha \simeq 0.0917$ would be intriguing. Then, $f^{(Z)} = f^{(A)} = e^2$.

If $f \neq f'$ distinctly, then another option $f'/f = (g'/g)^2 = \tan^2\theta_w$ might work, implying $f^{(Z)} = 2f \sin^2\theta_w \cot^2 2\theta_w$ and $f^{(A)} = 2f \sin^2\theta_w$, and hence $ f^{(A)}/f^{(Z)} = \tan^2 2\theta_w$. In addition, if $f = g^2$, then $f\sin^2\theta_w = e^2$ and so, $f^{(Z)} = 2e^2\cot^2 2\theta_w$ and $f^{(A)}  = 2e^2$.  

We can see that in the hypothetic new weak interaction (15) there is embedded the coupling

\begin{eqnarray} 
-\frac{1}{2}\sqrt{\!f^{(A)}}\, \varphi^{(A)} F^{\mu \nu} A_{\mu \nu}
\end{eqnarray}

\ni of electromagnetic field $F^{\mu \nu} = \partial^\mu A^{\nu} \!-\!  \partial^\nu A^{ \mu}$, displaying the electroweak symmetry spontaneously broken by the  Standard-Model Higgs mechanism (modified "\,$\!$photonic portal"\, to the hidden sector). In the formal limit of $\sin\theta_w \rightarrow 0$, the coupling (22) returns to its previous form (1) which defined originally the "\,$\!$photonic portal"\, (then $f'$ denotes $f$ and {\it vice versa}, while $\varphi$ is identified with the previous steron field $\varphi$).

Note that now a nonzero vacuum expectation value $<\!\!\varphi^{(A)}\!\!>_{\rm vac} \neq 0$ can generate the sterino magnetic moment

\begin{equation} 
\mu_\psi \equiv \frac{\sqrt{f^{(A)} f\,}\, \zeta\! <\!\!\varphi^{(A)}\!\!>_{\rm vac}}{2M^2}\,,
\end{equation}

\ni if the coupling

\begin{equation} 
\frac{\sqrt{f^{(A)} f\,} \zeta\!<\!\!\varphi^{(A)}\!\!>_{\rm vac}}{2M^2}\,\bar\psi \sigma_{\mu\,\nu} \psi F^{\mu \nu}
\end{equation}

\ni is invoked, following approximately (for small momentum transfers and with $\varphi^{(A)}\: = \:\:{<\!\!\varphi^{(A)}\!\!>_{\rm vac}}$) from the interactions (2) and (22).

We  assume in our model that only the vacuum expectation value of Standard-Model neutral Higgs boson breaks spontaneously the electroweak symmetry. Then, it must be $<\!\!\varphi_3\!\!>_{\rm vac} =\! 0$ for nonsterile $\varphi_3$, while it can be that $<\!\!\varphi\!\!>_{\rm vac} \neq\! 0$ spontaneously for sterile $\varphi$, since $<\!\!\varphi\!\!>_{\rm vac} \neq\! 0$ does not violate this symmetry ($\varphi\,$ is sterile, as it gets zero weak isospin and zero weak hypercharge, so it displays no Standard-Model gauge charges). In such an option, where $\varphi^+, \varphi^-, \varphi_3$ are not an additional Higgs boson multiplet, $\sqrt{\!f^{(Z)}\,}<\!\!\varphi^{(Z)}\!\!>_{\rm vac}  = \sqrt{\!f'\,} \sin \theta_w\! <\!\varphi\!>_{\rm vac} \neq\! 0$ and $\sqrt{\!f^{(A)}\,}{<\!\!\varphi^{(A)}\!\!>_{\rm vac}} = \sqrt{\!f'\,} \cos\theta_w <\!\!\varphi\!\!>_{\rm vac} \neq\! 0$.

A nonzero vacuum expectation value $<\!\!\varphi\!\!>_{\rm vac} \neq 0$ (implying $<\!\!\varphi^{(Z)}\!\!>_{\rm vac} \neq 0$ and ${<\!\!\varphi^{(A)}\!\!>_{\rm vac}} \neq 0$) can be also used to generate spontaneously masses for nonsterile particles described by $\varphi^+, \varphi^-, \varphi_3$ as well as sterile particles corresponding to $\psi, \varphi, A_{\mu\,\nu}$ (in place of $\varphi_3$ and $\varphi$, the relevant physical fields are rather $\varphi^{(Z)}$ and $\varphi^{(A)}$).

In our model, the scalar fields $\varphi^+, \varphi^-, \varphi_3$ and $\varphi$ are assumed not to be coupled directly to the Standard-Model Higgs scalars (so, they cannot provide the "Higgs portal"\, to the hidden sector [2]). On the other hand, $\varphi^+, \varphi^-, \varphi_3$ can participate in the conventional Standard-Model electroweak interaction with the gauge bosons $W^+, W^-, W_3$ ($\varphi^+, \varphi^-, \varphi_3$ as well as $\varphi$ cannot be coupled in the \SM to the gauge boson $B$, since all of them get zero weak hypercharge). Thus, making use of the minimal gauge coupling, we can write the following electroweak interaction for $\varphi^+, \varphi^-, \varphi_3$:

\begin{equation} 
-g \sum_{k\,l\,m} \varphi_l(-i)\varepsilon_{k\,l\,m}i \partial_\mu \varphi_m W^\mu_k + \frac{1}{4} g^2 \sum_{k\,l} 
\sum_{m\,n\,p} \varphi_m(-i)\varepsilon_{k\,m\,n}(-i) \varepsilon_{l\,n\,p} \varphi_p W_{k\,\mu} W^\mu_l \,,
\end{equation}

\ni where $W^\mu_i$ and $\varphi_i$ are given as in Eqs. (13) and (14). Here, the weak isospin is described by three matrices $t_k = \left(-i\varepsilon_{k\,l\,m}\right)$. 

Then, after the spontaneous breaking of electroweak symmetry, we obtain for the first term in Eq. (25)

\begin{eqnarray} 
-g \sum_{k\,l\,m} \varphi_l(-i)\!\!\!&\!\!\!\!&\!\!\!\varepsilon_{k\,l\,m}i \partial_\mu \varphi_m W^\mu_k  \nonumber \\ &\!\!\!\!=\!\!\!\!& -g \left(\varphi^- i \partial_\mu \varphi_3 - \varphi_3 i \partial_\mu \varphi^-\right) W^{+ \mu} + 
g \left(\varphi^+ i \partial_\mu \varphi_3 - \varphi_3 i \partial_\mu \varphi^+\right) W^{- \mu} \nonumber \\  &\!\!\!\!\!\!\!\!& + g \cos\theta_w\left(\varphi^- i \partial_\mu \varphi^+ - \varphi^+ i \partial_\mu \varphi^-\right)\!\! Z^\mu - 
e \left(\varphi^- i \partial_\mu \varphi^+  - \varphi^+ i \partial_\mu \varphi^-\right)\!\! A^\mu , 
\end{eqnarray}

\ni where $g = e/\sin\theta_w$ and $ g \cos\theta_w = e\cot\theta_w$, while $\varphi_3 = \sqrt{f^{(Z)}/f\,} \cos\theta_w \varphi^{(Z)}- \sqrt{f^{(A)}/f\,} \sin\theta_w \varphi^{(A)}$ due to Eqs. (18). Here, the electric current of $\varphi^+$ and $\varphi^-$ is $j^{(\varphi)}_\mu =e(\varphi^- i \partial_\mu \varphi^+  - \varphi^+ i \partial_\mu \varphi^-)$.

Note finally that due to the interaction (15) the nonsterile scalar bosons $\varphi^\pm$ and $\varphi^{(Z)}_{\rm ph}$ are likely to be unstable, decaying into $W^\pm$ and $Z$, respectively, and a virtual $A$ boson that in turn transits into a real photon due to the coupling (22) (with $\varphi^{(A)} = <\!\!\varphi^{(A)}\!\!>_{\rm vac} \neq 0$):  

\begin{equation} 
\varphi^\pm \rightarrow W^\pm A^* \rightarrow W^\pm \gamma \;\;,\;\; \varphi^{(Z)}_{\rm ph}\rightarrow Z A^* \rightarrow Z \gamma
\end{equation}

\vspace{0.2cm}

\ni (if $m_{\varphi^\pm} > M_W$ and $m_{\varphi^{(Z)}} > M_Z$). Similarly, for the nonsterile scalar boson $\varphi^{(A)}_{\rm ph}$ we infer from the interaction (15) and the coupling (22) (with $\varphi^{(A)} = <\!\!\varphi^{(A)}\!\!>_{\rm vac} \neq 0$)  that 

\begin{equation} 
\varphi^{(A)}_{\rm ph} \rightarrow \gamma A^* \rightarrow \gamma \gamma
\end{equation}

\vspace{0.2cm}

\ni (always $m_{\varphi^{(A)}} >0$), what makes $\varphi^{(A)}_{\rm ph}$ unstable. Here, generically, $\varphi^{(Z,A)} =  <\!\!\varphi^{(Z,A)}\!\!>_{\rm vac}+ \varphi^{(Z,A)}_{\rm ph} $.  

In contrast to the neutral scalar bosons described by $\varphi^{(Z)}$ and $\varphi^{(A)}$ (involving steron field $\varphi$ in their wave structure), sterinos corresponding to $\psi$ are in our model always stable (at least, their lowest generation, if there are more than one). They may constitute the cold dark matter as their thermal relic. They can annihilate in sterino-antisterino pairs, for instance, as $\bar{\psi} \psi \rightarrow A^* \rightarrow \gamma^* \rightarrow e^+ e^-$. The sterile mediating bosons $A$ are here unstable, decaying, for example, as $A \rightarrow \gamma^* \rightarrow e^+ e^-$ [1] (always $M > 2m_e$) and also as $A \rightarrow \varphi^{(A)}_{\rm ph}\gamma$ and $A \rightarrow \bar{\psi}\psi$, if $M > m_{\varphi^{(A)}}$ and $M> 2m_\psi$, respectively. They can be produced, in particular, in the inelastic Compton scattering $ e^- \gamma \rightarrow e^- \gamma^* \rightarrow e^- A$ at high energies.

\vspace{0.4cm}

\ni {\bf 3. Conclusion}

\vspace{0.4cm}

In the model of hidden sector presented in this note, in addition to the conventional Standard-Model electroweak interaction, there appears a hypothetic new weak interaction between hidden and Standard-Model sectors of the Universe, preserving primarily the electroweak symmetry and violating it eventually. After the spontaneous breaking of this symmetry by the Standard-Model Higgs mechanism, the new interaction realizes a "photonic portal"\, to the hidden sector of the Universe, embedded in the spontaneously broken electroweak symmetry. 

Since photon coupling dominates in effective electroweak interactions in the Standard-Model sector, the "photonic portal"\, introduced in our model dominates in effective interactions between hidden and Standard-Model sectors. This portal is narrow because of the large mass scale $M$ of sterile mediating bosons $A$.

Nowadays, more and more contents of particle physics come from particle astrophysics and cosmology that in a natural way are expected to be interested in the hypothetic hidden sector of the Universe as an option ({\it e.g.} Weinberg's recent book [3] illustrates the growing role of cosmology in modern physics). Since astrophysical observations are predominantly based on the detection of cosmic electromagnetic radiation and cosmic rays of various sorts (with a minor, though exciting, role played by recent direct-detection experiments on cold dark matter), the existence of a "photonic portal"\, to the hidden sector, conjectured in a previous work [1] and developed in the present note, may be an attractive hypothesis.

\vspace{0.6cm}
 

\vspace{0.4cm}

{\centerline{\bf References}}

\vspace{0.4cm}

\baselineskip 0.73cm 

{\everypar={\hangindent=0.65truecm}
\parindent=0pt\frenchspacing

{\everypar={\hangindent=0.65truecm}
\parindent=0pt\frenchspacing

~[1]~W.~Kr\'{o}likowski, {\it Acta Phys. Polon.} {\bf B 39}, 1881 (2008); arXiv: 0803.2977v2 [{\tt hep--ph}]; {\it Acta Phys. Polon.} {\bf B 40}, 111 (2009); arXiv: 0903.5163 [{\tt hep--ph}].

\vspace{0.2cm}

~[2]~{\it Cf. e.g.}  J. March-Russell, S.M. West, D. Cumberbath and D.~Hooper, {\it J. High Energy Phys.} {\bf 0807}, 058 (2008); K.~Kohri, J.~McDonald and N.~Sahu, arXiv: 0905.1312 [{\tt hep-ph}]; and references therein.  

\vspace{0.2cm}

~[3]~S.~Weinberg, {\it Cosmology}, Oxford University Press, New York, 2008.

\vfill\eject

\end{document}